\documentclass{epl}
\usepackage{amsmath,amsfonts,amssymb}
\usepackage[latin1]{inputenc}
\usepackage{graphicx} 
\usepackage{color} 

\title{Power law in the angular velocity distribution of a 
granular needle}

\author{J. Piasecki\inst{1} \and P. Viot\inst{2}}
\institute{\inst{1} Institute of Theoretical Physics, University of Warsaw,
 Ho\.za 69, 00 681 Warsaw, Poland\\
\inst{2} Laboratoire de Physique
Théorique de la Matière Condensée, Université Pierre et Marie Curie, 4, place
Jussieu, 75252 Paris Cedex, 05 France
}
\begin{document}
\maketitle
\begin{abstract}
We show how inelastic collisions induce a power law with exponent $-3$
in the decay  of   the angular velocity distribution  of   anisotropic
particles with sufficiently small   moment of inertia.  We investigate
this  question  within the Boltzmann  kinetic  theory for an elongated
granular particle immersed in a bath.  The  power law persists so long
as   the collisions are  inelastic   for  a large   range of   angular
velocities provided the mass ratio of the anisotropic particle and the
bath particles remains small.  Suggestions for observing this peculiar
feature are made.
\end{abstract}

Granular  gases are  systems  in  which  macroscopic particles  lose a
fraction of their  kinetic energy at  each collision \cite{BTE05,G05}.
In the  absence of energy  supply, starting from an homogeneous state,
the granular fluid  cools down\cite{H83} and after  a finite time  one
observes a    spontaneous  symmetry breaking  with the    formation of
clusters as well as of shear waves  and convection.  Conversely, if an
external energy  supply is continuously  brought to the particles, the
system   may reach   a   non-equilibrium steady  state   (NESS), whose
properties  differ significantly from   those of  thermal  equilibrium
(breakdown   of  the   equipartition\cite{MP99,BT02},     non-gaussian
statistics,  modified  hydrodynamics\cite{BDKS98},...).     All  those
characteristics  are intimately related to   the dissipative nature of
collisions.

The studies  of  granular   gases  have focused mainly    on spherical
particles\cite{PB03}.     Recent work\cite{GNB05}   on nearly   smooth
spherical grains stressed   the relevance  of  rotational degrees   of
freedom for the   hydrodynamic  behavior.  The rotational   energy  is
ubiquitous for anisotropic  particles,  and one can  reasonably expect
that the anisotropy (generally present in real granular systems) might
introduce additional effects.  Only  few studies have addressed  up to
now this question (free cooling state in a three-dimensional system of
needles \cite{HAZ99}, breakdown of the equipartition between different
degrees of freedom
\cite{VT04,GTV05}.  Experimentally, the dynamics of shaken granular dimer gas
has been investigated\cite{A005}.

In   this   letter, we  show  that  the   stationary  angular velocity
distribution of a  thin granular  needle of   mass $M$  acted upon  by
inelastic collisions in a  bath of point masses  $m$ shows a power-law
decay  in a large  range of  angular  velocities provided $M/m\ll 1$. We
relate this  feature to the properties  of a rigorous  solution of the
Boltzmann equation  in the $M/m\to  0$ limit.The region $M/m\ll 1$ studied
here is opposite to the Brownian motion case  $M/m\gg 1$. The latter has
been thoroughly  studied for  granular  fluids at least  for spherical
particles\cite{DG01, DB05}.  We discovered that the dynamics of a low
mass anisotropic  particle  is qualitatively different.  It  turns out
that  when the ratio $M/m$ is  finite but small  this unphysical (zero
moment of  inertia) solution continues  to  manifest itself through  a
power-law   decay  observed in a  large    range of angular velocities
whereas  a gaussian-like  decay  is restored  only for the high-energy
tail. When the restitution  coefficient is set  to $1$,  this peculiar
behavior disappears. Clearly, in real systems collisions with the ends
of the  elongated particle should  be  also considered.   The analysis
performed in  Ref.\cite{GTV05} has shown  that when the  length of the
particle  is large compared  to its width and  also to the size of the
bath particles the collisional  contributions from the extremities are
negligible and predictions  of  the simplified model considered   here
should apply.  We also   present suggestions for an experiment   which
would    permit   to observe  the    power-law   decay of  the angular
distribution function.

Quite recently a class of stationary states with power-law high-energy
tails (implying an infinite energy) has been found for homogeneous 
granular systems with the suggestion of implementing them by 
energy injection at large scales only  \cite{BM05,BMM05}. It should be
stressed however, that the underlying mechanism is completely
different from the one found here.  We  deal with an impurity
(the needle) immersed  in the granular fluid (which is
in a NESS). The power law appears  owing to the collisional rescaling
of the  distribution  of the bath  with  the weight  depending on  the
position of the impact point.  The anisotropy of the impurity plays in
this rescaling an important role.  The possibility  of a power law for
an impurity has been already mentioned 
in the review on the Maxwell model \cite{BK03}. However, to
get this effect the same power law had to be  assumed for the granular
bath.    Our  results reveal an   essentially   new  mechanism  for  a
physically relevant case of a free motion between collisions.

We consider a two-dimensional system  consisting of an infinitely thin
homogeneous needle of length  $L$ and moment of inertia $I=ML^{2}/12$,
with the fixed  center  of mass.    The needle undergoes   inelastic
collisions with the bath particles.    Between collisions it   rotates
freely around the axis passing through its center and perpendicular to
the plane of motion.  The only degree of freedom  of the needle is its
orientation specified by a unit vector ${\bf u}$ that points along its
axis.  The  rate  of change of  the  orientation $\dot{{\bf u}}=\omega {\bf
u}_\perp$  involves the angular velocity $\omega \in  ]-\infty,+\infty [$ and a unit vector
${\bf u}_\perp$ perpendicular to ${\bf u}$.

\begin{figure}[t]
\onefigure[width=10cm]{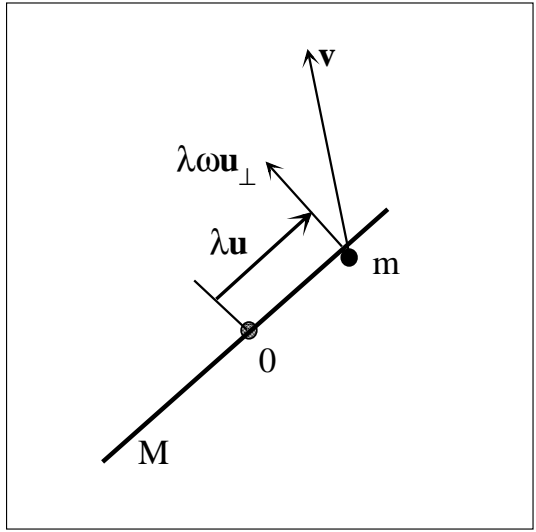}

\caption{The needle and a bath particle in the plane: 
${\bf u}$ is a unit vector along  the axis of  the needle forming with
${\bf u}^\perp$ an  orthonormal system.  The element of  the  needle at $\lambda
{\bf u}$  moves with  linear velocity ${\lambda\omega\bf   u}^\perp$. For a point  of
impact $\lambda {\bf u}$ the  pre-collisional relative velocity equals ${\bf
V}={\bf v}-\lambda \omega {\bf u}_\perp$}
\label{fig:1}
\end{figure}

At a binary collision the needle is hit by a point mass of the bath at
some point  $\lambda  {\bf  u},\;\;  |\lambda|<L/2$  where  $\lambda$  is the  algebraic
abscissa  along  the needle   (see  Fig.\ref{fig:1}).   The   relative
velocity ${\bf V}$ at the point of impact equals
\begin{equation}\label{eq:1}
{\bf V}={\bf v}-\lambda \dot{\bf u} = {\bf v}-\lambda \omega {\bf u}_\perp
\end{equation}
where ${\bf v}$ denotes the velocity of the bath particle.

The   instantaneous  binary  collisions  conserve   the total  angular
momentum.  Labelling the post-collisional  quantities  with a star and
using for any vector ${\bf w}$ the convenient notation $w_\perp = {\bf w}\cdot
{\bf u}_\perp$, $w_{||} = {\bf w}\cdot {\bf u}$ we thus write
\begin{equation}\label{eq:2}
I\omega^* + \lambda m v^*_\perp = I\omega + \lambda m v_\perp
\end{equation}
The  relative velocity will be assumed to change
according to the collision law
\begin{align}\label{eq:3}
V^*_\perp &=-\alpha V_\perp \\
V^*_{||} &= V_{||} \label{eq:4}
\end{align}
involving  the normal  restitution coefficient $0\leq\alpha \leq 1$.

By  combining  Eqs.~(\ref{eq:1})-(\ref{eq:4}),  we find the collisional
change of the particle velocity

\begin{equation}\label{eq:5}
v^*_\perp= v_\perp  -\frac{I(1+\alpha)V_\perp }
{I+ m\lambda^2}.
\end{equation} 
and the corresponding change of angular velocity 
\begin{equation} \label{eq:6}
\omega^*=\omega +\frac{(1+\alpha)V_\perp m^2\lambda }{I+m\lambda^2}.
\end{equation}
The inverse transformation is obtained by replacing $\alpha$ by $\alpha^{-1}$.

The   stationary   Boltzmann   equation  for   the  angular   velocity
distribution  $F(\omega)$ of the  needle  expresses the  invariance  of the
distribution under collisional   processes.  The gain term  corresponding to 
the  post-collisional angular velocity   $\omega$ must be
exactly compensated by  the  loss term involving the angular velocity 
$\omega$ as the pre-collisional one. The equation reads
\begin{align}\label{eq:7}
 \int_{-L/2}^{L/2} \!\!d\lambda &\int\!    d{\bf v} |v_\perp-\lambda \omega|
 \left(\frac{F(\omega^{**})\Phi_B({\bf v}^{**})}{\alpha^2}-F(\omega)
\Phi_B({\bf v})\right)=0,
\end{align} 
where the  pre-collisionnal  velocities  ${\bf v}^{**}$   and $\omega^{**}$
follow from  Eqs.(\ref{eq:2})-(\ref{eq:6})  by switching the  roles of
initial and final velocities and replacing $\alpha$ by $\alpha^{-1}$.  $\Phi_B({\bf
v})$ is  a velocity distribution describing   the steady state  of the
bath.
 
The integration over the velocity component $v_{||}$ in Eq.(\ref{eq:7}) can 
be readily performed as this variable does not show in the arguments of $F$.
Putting then  $v_\perp = (\lambda\omega) y$ 
one finds the integral equation

\begin{align}\label{eq:8}
&\int_{-L/2}^{L/2} d \lambda \lambda^2\int dy |y|F\left(\omega \left(1+y\frac{(1+\alpha)m\lambda^2}{I+m\lambda^2}\right)\right)
\phi_B\left(\lambda\omega \left(1+y\frac{(\alpha m\lambda^2-I)}{I+m\lambda^2}\right)\right)\nonumber\\&
= F(\omega)\int_{-L/2}^{L/2} d \lambda \lambda^2\int dy |y-1 |\phi_B(\lambda\omega y )
\end{align}
with $\phi_B(v)= \int dv_{||}\Phi_B(|{\bf v}|)$

We consider first the limit where both the mass of  the needle and the
coefficient  of  restitution $\alpha  $ tend  to zero.    In  this case the
velocity of  the bath particles is  not modified by collisions whereas
the  angular velocity of the  asymptotically  massless needle is reset
after each collision acquiring  instantaneously  the value $v_\perp   /\lambda$,
i.e. the ratio of the bath particle approach  velocity to the abscissa
of the impact point. The Boltzmann equation (\ref{eq:8}) becomes

\begin{figure}[t]

\onefigure[width=10cm]{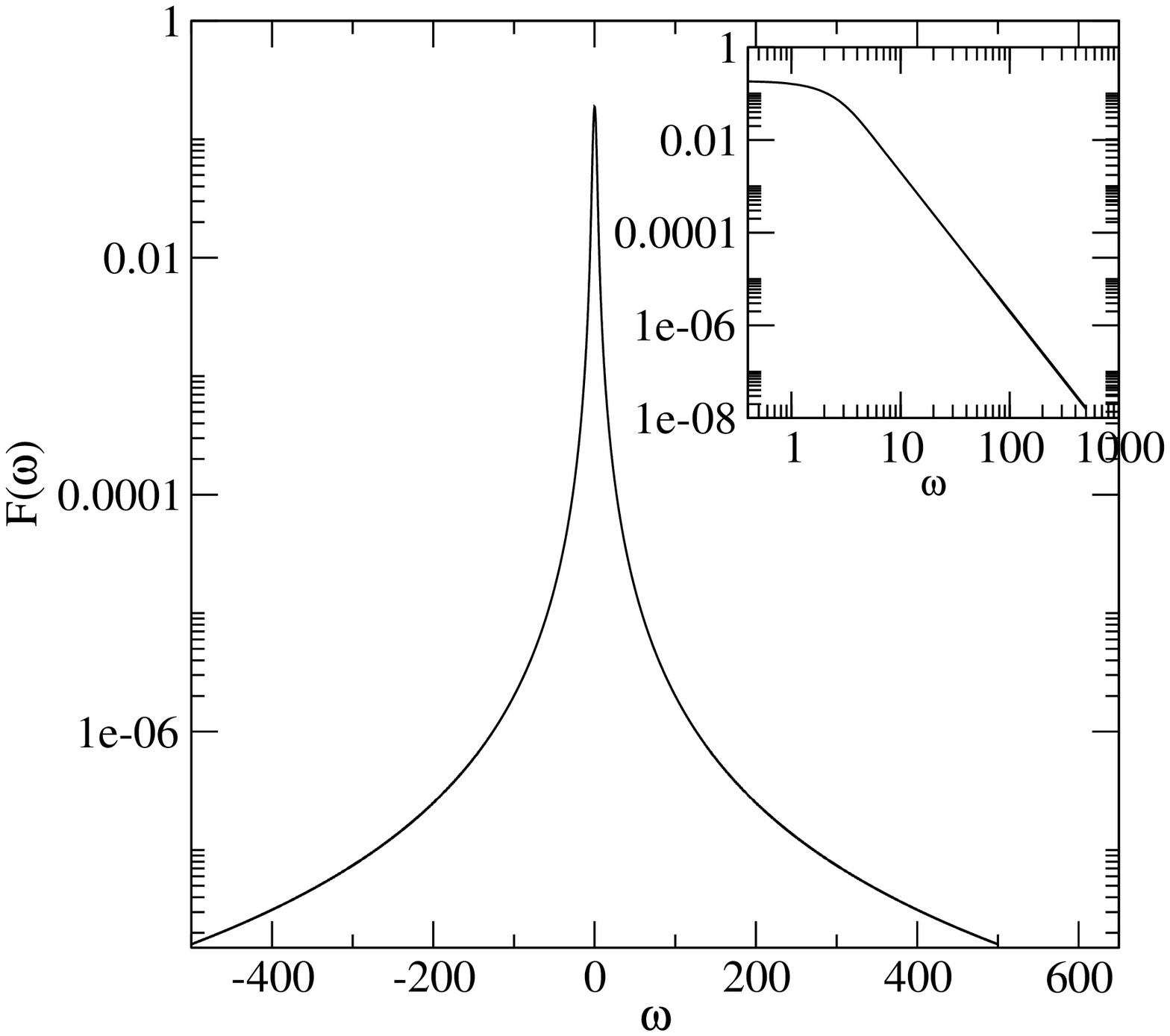}
\caption{Log-linear plot $F(\omega)$ of the ``zero mass''  
needle. The  inset  displays the log-log  plot  showing  the power-law
decay for sufficiently large $\omega $ values for $T/(mL^2)=1/2$.}\label{fig:2}
\end{figure}

\begin{align}\label{eq:9}
&\int_{-L/2}^{L/2} d \lambda \lambda^2\int dy |y|F\left(\omega \left(1+y\right)\right)
\phi_B\left(\lambda\omega\right)
= F(\omega)\int d \lambda \lambda^2\int dy |y-1 |\phi_B(\lambda\omega y )
\end{align}
The exact solution $F(\omega )$ of Eq.(\ref{eq:9}) is obtained by scaling 
the bath distribution $\phi_B(v)$ with the weight depending on the 
position of the impact point
\begin{equation}\label{eq:10}
 F(\omega)=\int_{-L/2}^{L/2} d\lambda \left(\frac{2\lambda}{L}\right)^{2}\phi_B(\lambda \omega )
\end{equation}
It should  be  stressed that this result   does  not require  specific
assumptions  about the distribution   of the bath  particles.  Indeed,
Eq.(\ref{eq:10}) implies the  power law decay $F(\omega)\propto \omega^{-3}$ for large
values of $\omega$ provided the second  moment of the distribution $\phi_B(v)$
exists.  The specific shape  of $\phi_B(v)$ is  irrelevant.  We also note
that for the rescaled distribution (\ref{eq:10}) the second and higher
moments   are not defined.   It is  thus   natural  to think that  the
stochastic variable $\omega$ is subject to Levy flights.

When the bath is at thermal equilibrium described by the Maxwell distribution 
$\phi_B(v)=\sqrt{m/2\pi T}\exp(-mv^2/2T),$ the stationary distribution takes
the form 
\begin{align}\label{eq:11}
F(\omega)&=-\sqrt{\frac{T}{2\pi m}}\frac{4}{L\omega^2}\exp\left(-\frac{mL^2\omega^2}{8T}\right)
\nonumber\\&+
\frac{4T}{L^{2}m\omega^3} {\rm erf}\left(\sqrt{\frac{mL^2}{8T}}\omega \right)
\end{align}
The power-law is reached for  $\omega >\sqrt{8T/(mL^2)}$.  Fig.~\ref{fig:2}
shows the logarithm of  the distribution  function versus the  angular
velocity. The inset   in Fig.\ref{fig:2}  displays  the log-log   plot
illustrating the crossover to the power-law behavior.

\begin{figure}[t]
\onefigure[width=10cm]{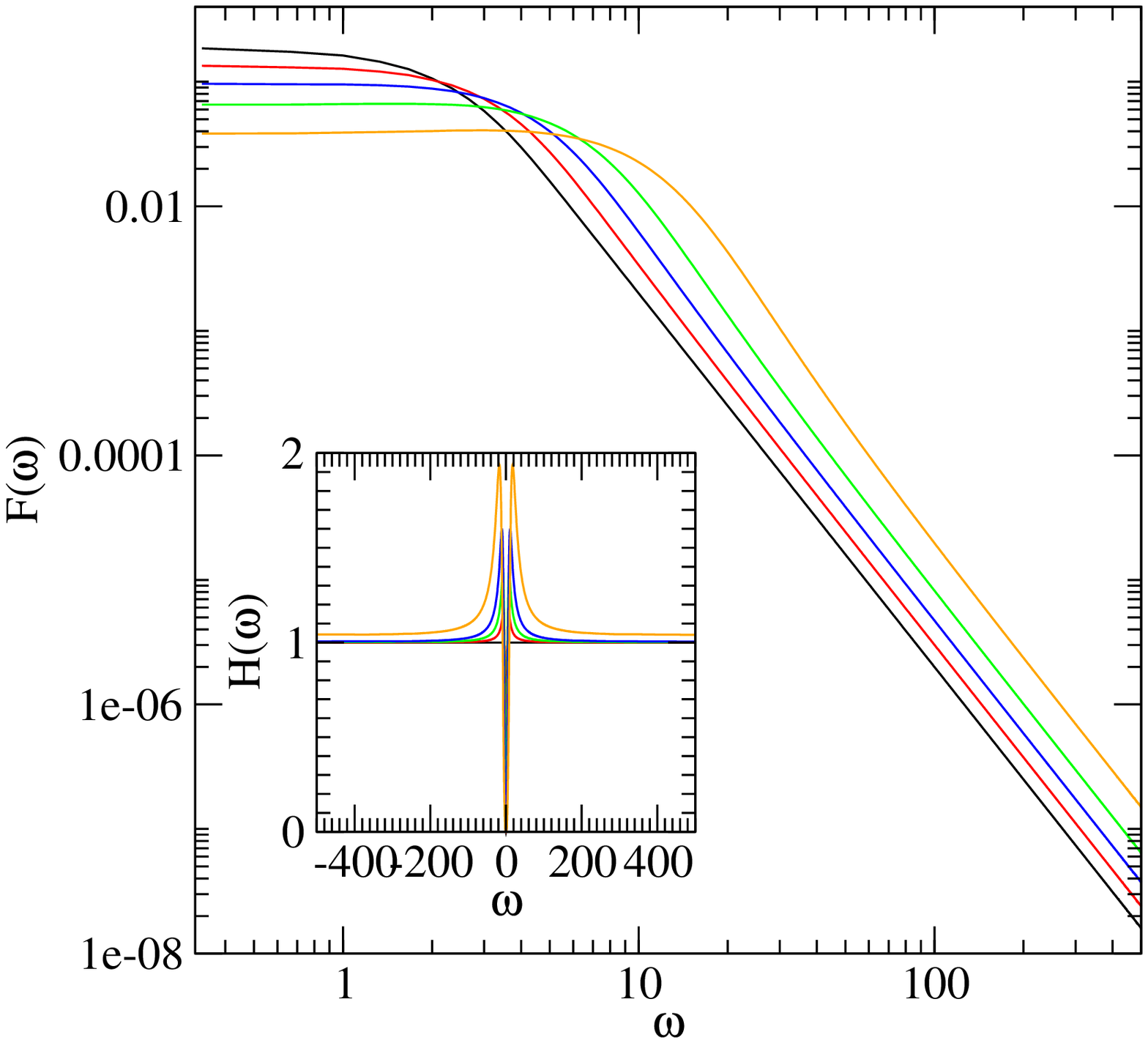}
\caption{Log-Log plot of $F(\omega)$  of the ``zero mass''  needle with different values 
of the coefficient  of  restitution $\alpha=0,0.2,0.4,0.4,0.8$ from  top to
bottom.  The inset shows  the $\omega$-dependence of the  rescaled function
$H(\omega)=F(\omega)\frac{mL^2(1-\alpha)}{4T(1+\alpha)} \omega^{3}$  for $T/(mL^2)=1/2$.}\label{fig:3}
\end{figure}

Eqs.(\ref{eq:10},\ref{eq:11})  correspond to  the maximal inelasticity
$\alpha =0$.  When $M=0$  and $0 < \alpha  < 1$, the  solution of the  Boltzmann
equation   behaves   asymptotically         as 
\begin{equation}
F(\omega)\sim \frac{4T(1+\alpha)}{mL^2(1-\alpha)\omega^{3}}.
\end{equation}
 Direct numerical integration of Eq.(\ref{eq:8}) (with $I=0$),
along the  same lines as in  \cite{BMP02}, is shown in Fig.\ref{fig:3}
for various  values  of $\alpha$.   The  inset of  Fig.\ref{fig:3} displays
$H(\omega)=F(\omega)\frac{mL^2(1-\alpha)}{4T(1+\alpha)} \omega^{3}$ versus $\omega$, indicating that
the asymptotic behavior is rapidly reached.  It should be noticed that
for elastic collisions ($\alpha =1$) all these effects disappear as then
\begin{equation}\label{eq:12}
F(\omega)= \sqrt{\frac{I}{2\pi T}}\exp\left(-\frac{I\omega^2}{2 T}\right)
\end{equation}
and the limit $I\to 0$ does not yield a probability distribution.

We turn now to the physically  relevant case of a  needle with a small
but finite mass $0< M \ll m$. An asymptotic analysis of the distribution
$F(\omega)$   can  be  then performed  on    the basis of  Eq.(\ref{eq:8}).
Assuming that at large angular velocities the distribution function is
gaussian, $F(w)\sim
\exp(-I\omega^2/(2\overline{T})$  (an assertion well supported by numerical
results\cite{PTV05}),  one  can   show that    $\overline{T}=(1+\alpha)T/2$,
irrespective of  the mass ratio of the  needle and the  bath particles
(it is an exact asymptotics in  the Brownian motion limit $M/m\to\infty$). By
means of  an  accurate numerical solution   of the  Boltzmann equation
(details will be  given elsewhere \cite{PTV05}), we  could essentially
confirm this asymptotic behavior inferred first by analytic arguments.
In addition, in the case under consideration ($0<  M/m \ll 1$), we could
identify a sub-leading  multiplicative term (decreasing  less rapidlly
than  a gaussian) which depends  on the mass ratio as   well as on the
coefficient of  restitution.   This combined behavior occurs   in  the
asymptotic regime.

\begin{figure}[t]
\onefigure[width=10cm]{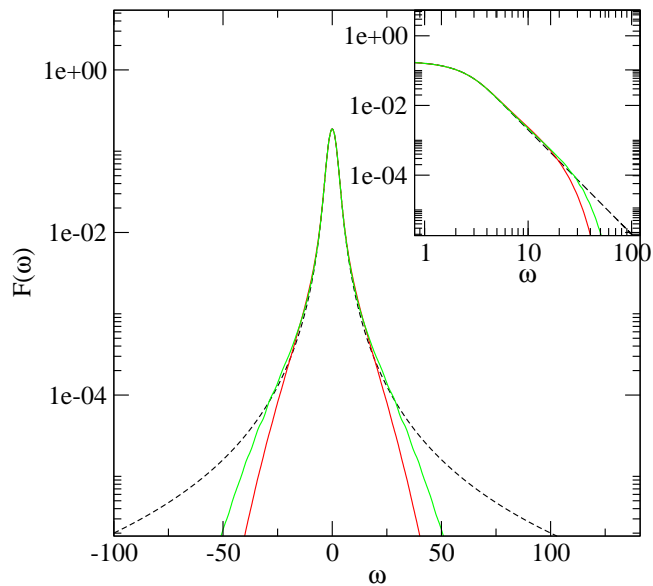}
\caption{Log-Log plot of $F(\omega)$ for a needle with a mass  $M/m=0.005,0.01$.(dashed curve corresponds to $M=0.0$). The inset shows the crossover between the power law and the gaussian-like asymptotics}\label{fig:4}
\end{figure}

However, our really important conclusion  is that before the very high
energy asymptotics is attained  the  power-law $\sim \omega^{-3}$ found for  a
massless   needle remains present with    a crossover occuring only at
$\omega_c\sim\sqrt{2T/I}$.  This  situation  is illustrated in Fig~\ref{fig:4}
where  the log-linear plot   of the angular distribution  functions is
displayed for mass ratios  $M/m=0,  0.005,0.01$, and the  dashed curve
represents the zero-mass   limit.    For $\omega<\omega_c$, one    observes  the
power-law  decay (see  the  log-log  plot in   the inset) whereas  the
gaussian behavior is   recovered beyond $\omega_c$.   Contrary to  the zero
mass case,  the existence of a gaussian  decay at very large values of
the velocity leads to a finite  granular temperature for a needle with
a   finite mass, and  more generally  all  moments of the distribution
functions  are now well   defined.  In fact,    one can show  that the
granular temperature defined as  the second moment of the distribution
times the moment of inertia  goes to zero with the  mass of the needle
if $\alpha <1$, whereas the temperature remains independent  of the mass if
$\alpha=1$. This  result is also  a consequence of  the existence of a well
defined solution of the Boltzmann equation for a zero-mass needle when
$\alpha<1$. In   conclusion,  the  behavior  for  $M/m\ll  1$ is qualitatively
different from that observed in the Brownian motion limit involving an
 new collisional scaling  mechanism producing the $\omega^{-3}$
power law.

Since  the power-law regime corresponds   to low and moderate  angular
velocities,   it  should    be    accessible   to    an   experimental
observation. Recent experiments on intensely vibrated granular systems
using  high  speed  photography\cite{FM02,FM04}, image        analysis
techniques and   particle tracking\cite{WHP01,WP02}   show  that  many
quantities  (granular temperature,  velocity   profiles,...)    can be
precisley measured   allowing  the   study of  the   equipartition  or
fluctuation theorem.  Several orders of magnitude of velocities can be
monitored\cite{A005,RM00} within  the region where the power-law decay
is expected.   More  specifically, by  immersing  in a two-dimensional
granular gas a thin rigid needle which  rotates around a perpendicular
axis, it   should   be  possible   to   obtain the  angular   velocity
distribution in the relevant region of $\omega$.  By using different needle
lengths or  different sizes of  the bath particles  one should observe
successive crossover  velocities.
We expect that the collisional rescaling found here will induce the
power law also in the case of anisotropic particles of a general convex 
shape in the region accessible to experiment.

J. P. acknowledges  financial support by the CNRS, France, and
the hospitality at the {\it Laboratoire de Physique Théorique de la Matière
Condensée, UPMC  (Paris)} where this research 
has been carried out.


\begin{thebibliography}{10}

\bibitem{BTE05}
A. Barrat, E. Trizac, and M.~H. Ernst, J. Phys. Condens.Matter {\bf 17},  S2429
   (2005).

\bibitem{G05}
I. Goldhirsch, Annu. Rev. Fluid. Mech. {\bf 35},  267  (2003).

\bibitem{H83}
P.~K. Haff, J. Fluid Mech. {\bf 134},  401  (1983).

\bibitem{MP99}
P.~A. Martin and J. Piasecki, Europhys. Lett {\bf 46},  613  (1999).

\bibitem{BT02}
A. Barrat and E. Trizac, Granul. Matter {\bf 4},  57  (2002).

\bibitem{BDKS98}
J.~J. Brey, J.~W. Dufty, C.~S. Kim, and A. Santos, Phys. Rev. E {\bf 58},  4638
   (1998).

\bibitem{PB03}
T. P\"oschel and N. Brilliantov, {\em Granular Gas Dynamics} (Springer, Berlin,
  2003).

\bibitem{GNB05}
I. Goldhirsch, S.~H. Noskowicz, and O. Bar-Lev, Phys. Rev. Lett. {\bf 95},
  068002  (2005).

\bibitem{HAZ99}
M. Huthmann, T. Aspelmeier, and A. Zippelius, Phys. Rev. E {\bf 60},  654
  (1999).

\bibitem{VT04}
P. Viot and J. Talbot, Phys. Rev. E {\bf 69},  051106  (2004).

\bibitem{GTV05}
H. Gomart, J. Talbot, and P. Viot, Phys. Rev. E {\bf 71},  051306  (2005).

\bibitem{A005}
J. Atwell and J.~S. Olafsen, Phys. Rev. E {\bf 71},  062301  (2005).

\bibitem{DG01}
J.~W. Dufty and V. Garzo, J. Stat. Phys. {\bf 105},  723  (2001).

\bibitem{DB05}
J.~W. Dufty and J.~J. Brey, New Journal of Physics {\bf 7},  20  (2005).

\bibitem{BM05}
E. Ben-Naim and J. Machta, Phys. Rev. Lett. {\bf 94},  138001  (2005).

\bibitem{BMM05}
E. Ben-Naim, B. Machta, and J. Machta, Phys. Rev. E {\bf 72},  021302  (2005).

\bibitem{BK03}
E. Ben-Naim and P. Krapivski,  in {\em Granular Gas Dynamics} (Springer,
  Berlin, 2003), p.\ 64.

\bibitem{BMP02}
T. Biben, P.~A. Martin, and J. Piasecki, Physica A {\bf 310},  308  (2002).

\bibitem{PTV05}
J. Talbot, J. Piasecki, and P. Viot, unpublished (unpublished).

\bibitem{FM02}
K. Feitosa and N. Menon, Phys. Rev. Lett. {\bf 88},  198301  (2002).

\bibitem{FM04}
K. Feitosa and N. Menon, Phys. Rev. Lett.  164301  (2004).

\bibitem{WHP01}
R.~D. Wildman, J.~M. Huntley, and D.~J. Parker, Phys. Rev. Lett. {\bf 86},
  3304  (2001).

\bibitem{WP02}
R.~D. Wildman and D.~J. Parker, Phys. Rev. Lett. {\bf 88},  064301  (2002).

\bibitem{RM00}
F. Rouyer and N. Menon, Phys. Rev. Lett. {\bf 85},  3676  (2000).

\end{thebibliography}
\end{document}